# Homogenization of a Biot-Stokes system modeling deformable vuggy porous media

Zhaoqin Huang[1], Xu Zhou[1], Tao Huang[1], Jun Yao[1*], Xiaoguang Wang[2*], Hervé Jourde[2]

1. Research Center of Multiphase Flow in Porous Media, China University of Petroleum (East China), Qingdao Shandong, 266580, China

2. Laboratoire HydroSciences Montpellier (HSM), UMR 5569 CNRS-IRD-UM, Université de Montpellier, 163 Rue Auguste Broussonnet, Montpellier 34095, France

*Corresponding author: yaojunhdpu@126.com, xiaoguang.wang@umontpellier.fr

## Abstract

Vugs are small to medium-sized cavities commonly present in rocks. The presence of vugs may have non-trivial impacts on the hydro-mechanical behaviors of the rock. How to effectively quantify and analyze such effects is still an challenging problem. To address the problem, we derive a macroscopic coupled hydro-mechanical model for single-phase viscous fluid flow through a deformable vuggy porous medium. We first model the hydro-mechanical coupling process on a fine scale using Biot's equations within porous matrix, Stokes equations within the vugs, and an extended Beavers-Joseph-Saffman boundary condition on the porous-fluid interface. Based on the homogenization theory, we then obtain macroscopic Biot's equations that govern the hydro-mechanical coupling behaviors of vuggy porous media on a large scale. Subsequently, the macroscopic poroelastic coefficients, such as the effective permeability, effective Young's modulus and effective Biot coefficient, are derived from three cell problems. Finally, several numerical examples are designed to validate the proposed model and to demonstrate the computational procedure for evaluation of the hydro-mechanical behaviors of vuggy porous media.

## 1. Introduction

Vug are defined as visible pores that are significantly larger than adjacent grains or crystals (Lucia, 2007; Huang et al., 2010, 2011), as shown in Figure 1. The presence of vugs can significantly increase the porosity and permeability of the rock (Arbogast and Lehr, 2006; Popov et al., 2009; Karasuyama et al., 2011; Xie et al., 2017). The main challenge is the coupling of porous flow and free-fluid flow (Arbogast and Brunson, 2007; Zhang et al., 2016), which is the key issue for many environmental, biomedical and industrial applications (Mosthaf et al., 2011; Baber et al., 2012; Chen et al., 2014). There are usually two schemes to model the coupling porous-free flow. One is the single domain approach using Stokes-Brinkman equations, the other one is the two domain approach using Darcy-Stokes equations. The former provides a model that can be continuously varied from a Darcy dominated flow to a Stokes dominated flow, so it avoids to introduce extra boundary conditions between porous media and free fluid region (Popov et al., 2009). However, the recent results show that the Stokes-Brinkman equations are only applicable for a special parameter set (Karasuyama et al., 2011), where the effective vugs permeability should be less than four orders of magnitude different from the matrix permeability. Therefore, the Darcy-Stokes system is more used. But one should introduce suitable boundary conditions at the porous-fluid interface (Goyeau et al., 2003) to couple these two equations. The Beavers-Joseph-Saffman (BJS) boundary condition is usually adopted (Beavers and Joseph, 1967; Saffman, 1971; Mikelic and Jäger, 2000; Carraro et al., 2015). It should be note that the mentioned coupling equations are just focus on the fluid flow and ignore the deformation of the porous solid skeleton.

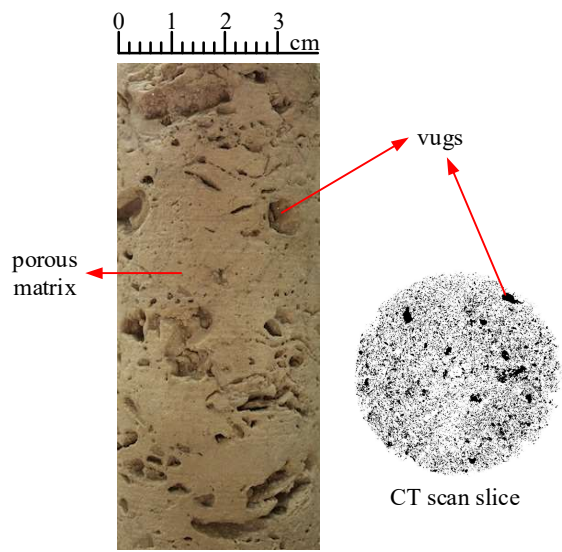

**Figure 1: A typical core of vuggy carbonate reservoirs.**

Recently, the interaction between the free fluid and a deformable porous medium has been attracted extensive attention because of its wide range of applications, e.g. ground surface water flow, seabed-wave, blood-vessel interactions and geomechanics of fractured vuggy carbonate reservoirs (Wang et al., 2004; Badia et al., 2009; Zhang et al., 2017). Let us focus on the last application. As illustrated in Figure 1, a vuggy carbonate reservoir is a typical double-porosity medium. Due to the different sizes between matrix pores and vugs, the vuggy porous media are more stress-sensitive than the homogeneous porous media (Jaeger et al., 2007; Zhang et al., 2017). The Biot's equations are usually used to model the hydro-mechanical coupling effects of porous media (Biot, 1941, 1956; Jaeger et al., 2007; Both et al., 2017).

Murad et al. (2000, 2001) were the first to use Biot-Stokes equations to study the hydro-mechanical behavior of a double-porosity medium, in which a porous cell contains micro pores and the surrounding system of macro pores, void spaces or bulk flow paths (e.g., fractured rock or aggregated soil). And the extended BJS boundary conditions were proposed in the framework of small strains. Following their works, Showalter (2005) proposed a mixed formulation for the resolvent Biot-Stokes equations with extended BJS boundary conditions. Whereafter, some numerical studies of the Biot-Stokes coupling problem have been presented (Badia et al., 2009; Ambartsumyan et al., 2017). Although we can use numerical techniques to get accurate and detailed poroelastic and fluid behaviors on the fine scale (the scale of the periodic region), it will lead to a computationally expensive problem if one focuses on the fine-scale structure on the macroscale (Kouznetsova et al., 2001; Yalchin and Hou, 2009; E, 2011). Therefore, an upscaling description of the fine-scale structural behaviors is needed when we focuses on the macroscale problem (e.g., core analysis and reservoir simulation).

Usually, one can get the macroscale constitutive relations through the experiments. This approach, however, is less effective when dealing with complex coupling multi-physical process. An alternative is provided by volume averaging or homogenization techniques (Whitaker, 1999; Auriault et al., 2009), which allow the inclusion of the fine-scale descriptions within the larger scale problem. Murad et al. (2000, 2001) applied the homogenization technique to upscale the Biot-Stokes equations aiming to get the constitutive laws for the double-porosity media. However, they assumed the local Biot number $Bi \to \infty$, and then they neglected the interfacial resistance. As a result, the continuity condition of the tangential velocity on the interface is enforced instead of the BJS condition. Lewandowska and Auriault (2013) developed a macroscopic model of hydro-mechanical coupling for the case of a porous medium containing isolated cracks or vugs. The Biot-Stokes equations are applied in their work, in which the continuity conditions of the stress, fluid pressure and velocity are imposed at the porous-fluid interface. More recently, Wan and Eghbalian (2016a, 2016b) studied the hydro-mechanical description of fractured porous media based on micro poromechanics by using the mean-field theory and Mori-Tanaka homogenization scheme. They also used Biot-Stokes equations on the fine scale, but the continuity of the velocity and stress are directly applied instead of the BJS condition. Therefore, a general homogenized result of a Biot-Stokes systems coupling with BJS condition is still an opening problem.

In this work, we start from the Biot-Stokes equations coupling with the BJS condition on fine scale to develop a general macroscopic hydro-mechanical model for vuggy porous media via the asymptotic expansions homogenization theory (Sanchez-Palencia, 1980; Auriault et al., 2009; Argilaga et al., 2016). To our best knowledge, the derived macroscopic Biot's equations are novel and general for deformable vuggy porous media. It is shown that the general structure of Biot's equations is the same as in the case of homogeneous medium but the effective poroelastic coefficients are modified and have different physical meanings. In Section 2, the Biot-Stokes equations of the problem at the fine scale is presented.

In Section 3, the homogenization upscaling process is described in detail. Section 4 gives several numerical examples and discussions.

## 2. Biot-Stokes equations on the fine scale

The development in this section is similar to that in the papers (Chen et al., 2004; Arbogast and Lehr, 2006). Let $\Omega$ be a bounded deformable fractured or vuggy porous medium in $\mathbf{R}^d$, $d=2, 3$, as depicted in Figure 2. And domain $\Omega$ is a union of non-overlapping regions $\Omega_m$ and $\Omega_f$, which are poroelastic matrix and free fluid region respectively. Let $\Gamma_{fm}=\Omega_m \cap \Omega_f$. The Biot's equations in $\Omega_m$ and the Stokes equations in $\Omega_f$, and the BJS interface conditions at the porous-fluid interface $\Gamma_{fm}$ are stated as follows.

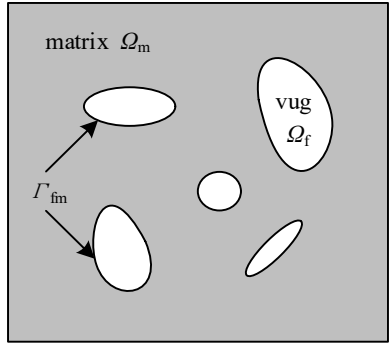

**Figure 2: Schematic of a periodic porous medium with micro vugs.**

(1) Biot's equations for poroelastic matrix

$$\begin{cases} \nabla \cdot \boldsymbol{\sigma}_m + \rho_m \boldsymbol{g} = 0 & \text{in } \Omega_m \\ \boldsymbol{\sigma}_m = -\boldsymbol{\alpha} p_p + \boldsymbol{\sigma}_s, \; \boldsymbol{\sigma}_s = \boldsymbol{a} : \boldsymbol{e}(\boldsymbol{u}_s) & \text{in } \Omega_m \\ \boldsymbol{v}_m = \boldsymbol{v}_p - \phi \dot{\boldsymbol{u}}_s = -\dfrac{\boldsymbol{k}}{\mu}\left(\nabla p_p - \rho_f \boldsymbol{g}\right) & \text{in } \Omega_m \\ \nabla \cdot \boldsymbol{v}_m = -\boldsymbol{\alpha} : \boldsymbol{e}(\dot{\boldsymbol{u}}_s) - \gamma \dot{p}_p & \text{in } \Omega_m \end{cases} \quad (1)$$

where, $\sigma_m$ and $\sigma_s$ are the total stress tensor of matrix and stress tensor of solid matrix skeleton, respectively. The density of saturated matrix $\rho_m = \phi \rho_f + (1-\phi)\rho_s$, here $\phi$ is the porosity of matrix, $\rho_f$ and $\rho_s$ are the density of fluid and solid matrix skeleton. $p_p$ is the matrix pore pressure; $\boldsymbol{\alpha}$ is the Biot coefficient, which is a second rank tensor; $\boldsymbol{a}$ is the fourth rank elastic tensor. $\boldsymbol{u}_s$ is the displacement of solid matrix skeleton, and $\dot{\boldsymbol{u}}_s$ denotes the differentiation of $\boldsymbol{u}_s$ with respect to time. $\boldsymbol{v}_m$ is the Darcy's velocity, and $\boldsymbol{v}_p$ is the average fluid velocity in matrix; $\boldsymbol{k}$ is the permeability tensor of matrix; $\mu$ is the fluid viscosity; $\boldsymbol{g}$ is the gravitation acceleration vector; $\gamma$ is the storage coefficient. The strain tensor $\boldsymbol{e}(\boldsymbol{u}_s)$ is defined by

$$\boldsymbol{e}(\boldsymbol{u}_s) = \dfrac{1}{2}\left(\nabla \boldsymbol{u}_s + \nabla^T \boldsymbol{u}_s\right) \quad (2)$$

(2) Stokes equations for free fluid

$$\begin{cases} \nabla \cdot \boldsymbol{\sigma}_f + \rho_f \boldsymbol{g} = 0 & \text{in } \Omega_f \\ \boldsymbol{\sigma}_f = -p_f \boldsymbol{I} + 2\mu \boldsymbol{e}(\boldsymbol{v}_f) & \text{in } \Omega_f \\ \nabla \cdot \boldsymbol{v}_f = 0 & \text{in } \Omega_f \end{cases} \quad (3)$$

where, $\sigma_f$ is fluid stress tensor; $p_f$ is the fluid pressure. $\boldsymbol{v}_f$ is free fluid velocity. The strain tensor $\boldsymbol{e}(\boldsymbol{v}_f)$ is

defined by

$$e(\boldsymbol{v}_f) = \frac{1}{2}\left(\nabla \boldsymbol{v}_f + \nabla^T \boldsymbol{v}_f\right) \tag{4}$$

(3) Interface conditions

$$\begin{cases} \boldsymbol{v}_f \cdot \boldsymbol{n} = (\boldsymbol{v}_d + \dot{\boldsymbol{u}}_s) \cdot \boldsymbol{n} & \text{on } \Gamma_{fm} \\ \boldsymbol{\sigma}_f \cdot \boldsymbol{n} = \boldsymbol{\sigma}_m \cdot \boldsymbol{n} & \text{on } \Gamma_{fm} \\ \boldsymbol{n} \cdot 2\mu e(\boldsymbol{v}_f) \cdot \boldsymbol{n} = p_f - p_p & \text{on } \Gamma_{fm} \\ \boldsymbol{n} \cdot 2\mu e(\boldsymbol{v}_f) \cdot \boldsymbol{\tau} = -\dfrac{\beta}{\sqrt{\boldsymbol{\tau} \cdot \boldsymbol{k} \cdot \boldsymbol{\tau}}}(\boldsymbol{v}_f - \dot{\boldsymbol{u}}_s) \cdot \boldsymbol{\tau} & \text{on } \Gamma_{fm} \end{cases} \tag{5}$$

where, $\boldsymbol{n}$ is the outward unit normal vector directed from $\Omega_f$ to $\Omega_m$; $\boldsymbol{\tau}$ is the unit tangential vector along the interface. The first equation is the mass balance, which requires that the normal fluid flux must be continuous across the interface. We note that this continuity of flux constrains the normal velocity of the solid matrix skeleton. The second equation represent the conservation of momentum, which requires that the stress of the porous matrix is balanced by the stress of the fluid. The third one is the balance of the normal components of the stress in the fluid phase across the interface. The fourth equation is the BJS condition modeling slip with friction, and $\beta$ is the BJS slip coefficient determined by experiments.

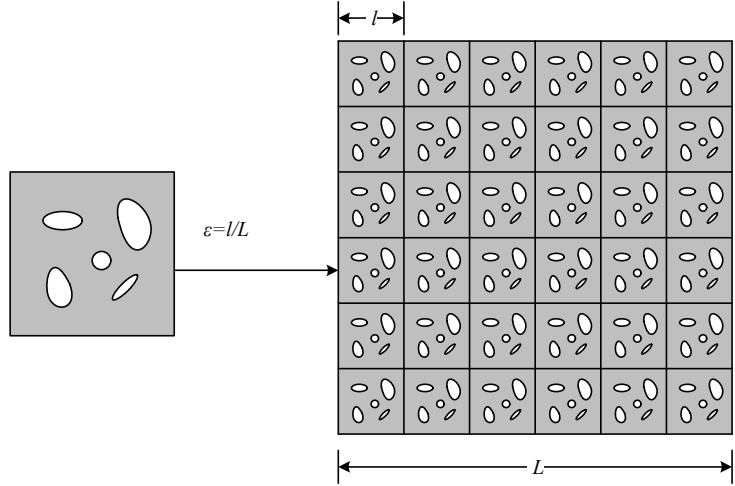

Figure 3: Schematic of a periodic vuggy porous medium

# 3. Upscaling based on homogenization theory

## 3.1 Preliminary and scaled problem

Now, we consider the homogenization procedure for the above Biot-Stokes system. The important characteristics of this procedure is existence of two different length scales: the fine scale $l$, which characterizes the typical REV (Representative Elementary Volume) size, and the macro scale $L$, which characterizes the global variation of external forces and boundary data. Let $\varepsilon = l/L$, and with $\varepsilon \ll 1$, as illustrated in Figure 3. Here, we also emphasize that the microscale $l_p \ll l$, which characterizes the matrix pore size. Let the vuggy porous medium $\Omega$ has a periodic mesostructured with period $Y$ (cf. Figure 3), where $Y = Y_\mathrm{m} \cup Y_\mathrm{f}$, with $Y_\mathrm{m}$ and $Y_\mathrm{f}$ being the matrix and free fluid parts, respectively. Define

$$\Omega_{\varepsilon\mathrm{m}} = \Omega \cap \{x : x \in \varepsilon Y_\mathrm{m}\}, \quad \Omega_{\varepsilon\mathrm{f}} = \Omega \cap \{x : x \in \varepsilon Y_\mathrm{f}\} \tag{6}$$

In this work, we only consider a formal asymptotic expansion of the solution in (1)~(5), and the boundary of $\Omega$ does not play a role in this expansion. So the above definition can be rewritten as follows

$$\Omega_{\varepsilon\mathrm{m}} = \{x : x \in \varepsilon Y_\mathrm{m}\}, \quad \Omega_{\varepsilon\mathrm{f}} = \{x : x \in \varepsilon Y_\mathrm{f}\} \tag{7}$$

Next, we consider the following scaled problem.

(1) Biot's equations in matrix

$$\begin{cases} \nabla \cdot \boldsymbol{\sigma}_\mathrm{m}^\varepsilon + \rho_\mathrm{m} \boldsymbol{g} = 0 & \text{in } \Omega_{\varepsilon\mathrm{m}} \\ \boldsymbol{\sigma}_\mathrm{m}^\varepsilon = -\alpha p_\mathrm{p}^\varepsilon + \boldsymbol{\sigma}_\mathrm{s}^\varepsilon, \quad \boldsymbol{\sigma}_\mathrm{s}^\varepsilon = \boldsymbol{a} : \boldsymbol{e}(\boldsymbol{u}_\mathrm{s}^\varepsilon) & \text{in } \Omega_{\varepsilon\mathrm{m}} \\ \boldsymbol{v}_\mathrm{m}^\varepsilon = \boldsymbol{v}_\mathrm{p}^\varepsilon - \phi \dot{\boldsymbol{u}}_\mathrm{s}^\varepsilon = -\dfrac{k^\varepsilon}{\mu}(\nabla p_\mathrm{p}^\varepsilon - \rho_\mathrm{f} \boldsymbol{g}) & \text{in } \Omega_{\varepsilon\mathrm{m}} \\ \nabla \cdot \boldsymbol{v}_\mathrm{m}^\varepsilon = -\boldsymbol{\alpha} : \boldsymbol{e}(\dot{\boldsymbol{u}}_\mathrm{s}^\varepsilon) - \gamma \dot{p}_\mathrm{p}^\varepsilon & \text{in } \Omega_{\varepsilon\mathrm{m}} \end{cases} \tag{8}$$

(2) Stokes equations in vugs

$$\begin{cases} \nabla \cdot \boldsymbol{\sigma}_\mathrm{f}^\varepsilon + \rho_\mathrm{f} \boldsymbol{g} = 0 & \text{in } \Omega_{\varepsilon\mathrm{f}} \\ \boldsymbol{\sigma}_\mathrm{f}^\varepsilon = -p_\mathrm{f}^\varepsilon \boldsymbol{I} + 2\mu\varepsilon^2 \boldsymbol{e}(\boldsymbol{v}_\mathrm{f}^\varepsilon) & \text{in } \Omega_{\varepsilon\mathrm{f}} \\ \nabla \cdot \boldsymbol{v}_\mathrm{f}^\varepsilon = 0 & \text{in } \Omega_{\varepsilon\mathrm{f}} \end{cases} \tag{9}$$

(3) Interface conditions

$$\begin{cases} \boldsymbol{v}_\mathrm{f}^\varepsilon \cdot \boldsymbol{n} = (\boldsymbol{v}_\mathrm{d}^\varepsilon + \dot{\boldsymbol{u}}_\mathrm{s}^\varepsilon) \cdot \boldsymbol{n} & \text{on } \Gamma_\mathrm{fm}^\varepsilon \\ \boldsymbol{\sigma}_\mathrm{f}^\varepsilon \cdot \boldsymbol{n} = \boldsymbol{\sigma}_\mathrm{m}^\varepsilon \cdot \boldsymbol{n} & \text{on } \Gamma_\mathrm{fm}^\varepsilon \\ \boldsymbol{n} \cdot 2\mu\varepsilon^2 \boldsymbol{e}(\boldsymbol{v}_\mathrm{f}^\varepsilon) \cdot \boldsymbol{n} = p_\mathrm{f}^\varepsilon - p_\mathrm{p}^\varepsilon & \text{on } \Gamma_\mathrm{fm}^\varepsilon \\ \boldsymbol{\tau} \cdot 2\mu \boldsymbol{e}(\boldsymbol{v}_\mathrm{f}^\varepsilon) \cdot \boldsymbol{n} = -\dfrac{\beta}{\varepsilon\sqrt{\boldsymbol{\tau} \cdot \boldsymbol{k}^\varepsilon \cdot \boldsymbol{\tau}}}(\boldsymbol{v}_\mathrm{f}^\varepsilon - \dot{\boldsymbol{u}}_\mathrm{s}^\varepsilon) \cdot \boldsymbol{\tau} & \text{on } \Gamma_\mathrm{fm}^\varepsilon \end{cases} \tag{10}$$

where, we have scaled both the fluid viscosity $\mu$ and permeability $k^\varepsilon$ by $\varepsilon^2$. This is the usual scaling for deriving Darcy's equation from Stokes equations (Arbogast and Lehr, 2006). Since as $\varepsilon \to 0$, flow paths (in our case matrix pores and vugs) will constrict, and a corresponding decrease in fluid viscosity is required to maintain the flow rates. Moreover, the permeability $\boldsymbol{k} = O(l_\mathrm{p}^2)$, so it have a similar scaling of viscosity. Note that these scaling treatments are implied in the stated scaling of BJS condition.

Following the custom of homogenization theory, we assume that any point is described by two coordinates: $\boldsymbol{x} \in \Omega$ describing the global location of the point and $\boldsymbol{y} \in Y$ describing the location of the point within the $\varepsilon$-cell $\varepsilon Y$. Consequently, $\boldsymbol{x}$ and $\boldsymbol{y}$ are related by the scaling constant $\varepsilon$: $\boldsymbol{y} \sim \varepsilon^{-1} \boldsymbol{x}$. By the chain rule, the following relation holds

$$\nabla = \nabla_x + \frac{1}{\varepsilon} \nabla_y \tag{11}$$

where $\nabla_x$ and $\nabla_y$ denotes the gradient operators with respect to $\boldsymbol{x}$ and $\boldsymbol{y}$, respectively. Assuming that the solution to (8)~(10) behave as if it was a function of these two coordinates and that it can be expanded in a power series in terms of $\varepsilon$, the stress, pressure, displacement and velocity are then can be expanded in the following asymptotic form

$$\varphi^{\varepsilon}(\boldsymbol{x},\boldsymbol{y},t)=\varphi^{(0)}(\boldsymbol{x},\boldsymbol{y},t)+\varepsilon\varphi^{(1)}(\boldsymbol{x},\boldsymbol{y},t)+\varepsilon^{2}\varphi^{(2)}(\boldsymbol{x},\boldsymbol{y},t)+\cdots \tag{12}$$

where, $\varphi^{(i)}$ denotes the different variables, e.g. stress, pressure, displacement and velocity, which are $Y$-periodic in $\boldsymbol{y}$, $\boldsymbol{x}\in\Omega$, $\boldsymbol{y}\in Y$. Then, we shall substitute (12) into (8)~(10), and analyze the resulting equations.

## 3.2 Two-scale asymptotic analysis

(1) Biot's equations in matrix

We now we shall substitute (12) into (8), apply (11), and collect terms with like powers of $\varepsilon^{-1}$ and $\varepsilon^{0}$, which yield

$$\begin{cases} \nabla_y \cdot \boldsymbol{\sigma}_m^{(0)} = 0 & \text{in } Y_m \\ \boldsymbol{\sigma}_m^{(0)} = -\alpha p_p^{(0)} + \boldsymbol{\sigma}_s^{(0)}, \quad \boldsymbol{a}:\boldsymbol{e}_y\left(\boldsymbol{u}_s^{(0)}\right)=0 & \text{in } Y_m \\ \boldsymbol{\sigma}_s^{(0)} = \boldsymbol{a}:\boldsymbol{e}_x\left(\boldsymbol{u}_s^{(0)}\right)+\boldsymbol{a}:\boldsymbol{e}_y\left(\boldsymbol{u}_s^{(1)}\right) & \text{in } Y_m \\ \boldsymbol{v}_m^{(0)} = \boldsymbol{v}_p^{(0)} - \phi\dot{\boldsymbol{u}}_s^{(0)} & \text{in } Y_m \\ \nabla_y p_p^{(0)} = 0 & \text{in } Y_m \\ \nabla_y \cdot \boldsymbol{v}_m^0 = -\boldsymbol{\alpha}:\boldsymbol{e}_y\left(\boldsymbol{u}_s^{(0)}\right)=0 & \text{in } Y_m \\ \nabla_x \cdot \boldsymbol{\sigma}_m^{(0)} + \nabla_y \cdot \boldsymbol{\sigma}_m^{(1)} + \rho_m \boldsymbol{g} = 0 & \text{in } Y_m \\ \mu \boldsymbol{k}^{-1}\boldsymbol{v}_m^{(0)} + \nabla_x p_p^{(0)} + \nabla_y p_p^{(1)} - \rho_f \boldsymbol{g} = 0 & \text{in } Y_m \\ \nabla_x \cdot \boldsymbol{v}_m^{(0)} + \nabla_y \cdot \boldsymbol{v}_m^{(1)} = -\boldsymbol{\alpha}:\left[\boldsymbol{e}_x\left(\dot{\boldsymbol{u}}_s^{(0)}\right)+\boldsymbol{e}_y\left(\dot{\boldsymbol{u}}_s^{(1)}\right)\right]-\gamma\dot{p}_p^{(0)} & \text{in } Y_m \end{cases} \tag{13}$$

By the second and fifth equations of (13), we have

$$\boldsymbol{u}_s^{(0)}=\boldsymbol{u}_s^{(0)}(\boldsymbol{x},t), \quad p_p^{(0)}=p_p^{(0)}(\boldsymbol{x},t) \tag{14}$$

That is, $\boldsymbol{u}_s^{(0)}$ and $p_p^{(0)}$ are independent of $\boldsymbol{y}$. This implies the intuition that the local average of $\boldsymbol{u}_s$ and $p_p$ does not oscillate.

(2) Stokes equations in vugs

$$\begin{cases} \nabla_y \cdot \boldsymbol{\sigma}_f^{(0)} = 0, \quad \boldsymbol{\sigma}_f^{(0)} = -p_f^{(0)}\boldsymbol{I} & \text{in } Y_f \\ \nabla_x \cdot \boldsymbol{\sigma}_f^{(0)} + \nabla_y \cdot \boldsymbol{\sigma}_f^{(1)} + \rho_f \boldsymbol{g} = 0 & \text{in } Y_f \\ \boldsymbol{\sigma}_f^{(1)} = -p_f^{(1)}\boldsymbol{I} + 2\mu\boldsymbol{e}_y\left(\boldsymbol{v}_f^{(0)}\right) & \text{in } Y_f \\ \nabla_y \cdot \boldsymbol{v}_f^{(0)} = 0 & \text{in } Y_f \\ \nabla_x \cdot \boldsymbol{v}_f^{(0)} + \nabla_y \cdot \boldsymbol{v}_f^{(1)} = 0 & \text{in } Y_f \end{cases} \tag{15}$$

By the first equation of (15), we have

$$p_{\rm f}^{(0)} = p_{\rm f}^{(0)}(\boldsymbol{x},t) \; , \; \boldsymbol{\sigma}_{\rm f}^{(0)} = \boldsymbol{\sigma}_{\rm f}^{(0)}(\boldsymbol{x},t) \tag{16}$$

So $\boldsymbol{\sigma}_{\rm f}^{(0)}$ and $p_{\rm f}^{(0)}$ are also independent of $\boldsymbol{y}$.

(3) Interface conditions

$$\begin{cases} \boldsymbol{v}_{\rm f}^{(0)} \cdot \boldsymbol{n} = \left(\boldsymbol{v}_{\rm m}^{(0)} + \dot{\boldsymbol{u}}_{\rm s}^{(0)}\right) \cdot \boldsymbol{n} & \text{on } Y_{\rm fm} \\ \boldsymbol{\sigma}_{\rm f}^{(0)} \cdot \boldsymbol{n} = \boldsymbol{\sigma}_{\rm m}^{(0)} \cdot \boldsymbol{n} & \text{on } Y_{\rm fm} \\ p_{\rm f}^{(0)} = p_{\rm p}^{(0)} & \text{on } Y_{\rm fm} \\ \boldsymbol{\sigma}_{\rm f}^{(1)} \cdot \boldsymbol{n} = \boldsymbol{\sigma}_{\rm m}^{(1)} \cdot \boldsymbol{n} & \text{on } Y_{\rm fm} \\ \boldsymbol{n} \cdot 2\mu \boldsymbol{e}_y\left(\boldsymbol{v}_{\rm f}^{(0)}\right) \cdot \boldsymbol{n} = p_{\rm f}^{(1)} - p_{\rm p}^{(1)} & \text{on } Y_{\rm fm} \\ \boldsymbol{\tau} \cdot 2\mu \boldsymbol{e}_y\left(\boldsymbol{v}_{\rm f}^{(0)}\right) \cdot \boldsymbol{n} = -\dfrac{\beta}{\sqrt{\boldsymbol{\tau} \cdot \boldsymbol{k} \cdot \boldsymbol{\tau}}}\left(\boldsymbol{v}_{\rm f}^{(0)} - \dot{\boldsymbol{u}}_{\rm s}^{(0)}\right) \cdot \boldsymbol{\tau} & \text{on } Y_{\rm fm} \\ \boldsymbol{v}_{\rm f}^{(1)} \cdot \boldsymbol{n} = \left(\boldsymbol{v}_{\rm m}^{(1)} + \dot{\boldsymbol{u}}_{\rm s}^{(1)}\right) \cdot \boldsymbol{n} & \text{on } Y_{\rm fm} \end{cases} \tag{17}$$

where $Y_{\rm fm} = Y_{\rm f} \cap Y_{\rm m}$, and the third equation implies that

$$p_{\rm f}^{(0)} = p_{\rm p}^{(0)} = p^{(0)}(\boldsymbol{x},t) \tag{18}$$

## 3.3 The analysis of mechanical behavior

Apply the first three equations of (13), the second equation of (13), the second equation of (17), and equation (18) to see that

$$\begin{cases} \nabla_y \cdot \left[\boldsymbol{a} : \boldsymbol{e}_y\left(\boldsymbol{u}_{\rm s}^{(1)}\right)\right] = 0 & \text{in } Y_{\rm m} \\ \left[\boldsymbol{a} : \boldsymbol{e}_y\left(\boldsymbol{u}_{\rm s}^{(1)}\right)\right] \cdot \boldsymbol{n} = \left[-\boldsymbol{a} : \boldsymbol{e}_x\left(\boldsymbol{u}_{\rm s}^{(0)}\right) - (\boldsymbol{I} - \boldsymbol{\alpha}) p^{(0)}\right] \cdot \boldsymbol{n} & \text{on } Y_{\rm fm} \end{cases} \tag{19}$$

This system forms an elliptic problem in $\boldsymbol{y}$ for $\boldsymbol{u}_{\rm s}^{(1)}$ that can be solved in terms of $\boldsymbol{e}_x\left(\boldsymbol{u}_{\rm s}^{(0)}\right)$ and $p^{(0)}$. Therefore, by applying the linearity property, $\boldsymbol{u}_{\rm s}^{(1)}$ can be represented in terms of $\boldsymbol{e}_x\left(\boldsymbol{u}_{\rm s}^{(0)}\right)$ and $p^{(0)}$

$$\boldsymbol{u}_{\rm s}^{(1)} = \boldsymbol{\xi}(\boldsymbol{y}) : \boldsymbol{e}_x\left(\boldsymbol{u}_{\rm s}^{(0)}(\boldsymbol{x},t)\right) + \boldsymbol{\zeta}(\boldsymbol{y}) \cdot (\boldsymbol{I} - \boldsymbol{\alpha}) p^{(0)}(\boldsymbol{x},t) \tag{20}$$

where, $\boldsymbol{\xi}(\boldsymbol{y}) = \xi_i^{kh}$ is a third rank tensor; $\boldsymbol{\zeta}(\boldsymbol{y})$ is a vector. Substitute (20) into (19), we can get the following two local base cell problems. First, for each $k$ and $h$ ($k, h = 1, 2, 3$), we define the vector $\boldsymbol{\xi}^{kh}(\boldsymbol{y})$ to be the solution of

$$\begin{cases} \nabla_y \cdot \left[\boldsymbol{a} : \boldsymbol{e}_y\left(\boldsymbol{\xi}^{kh}\right)\right] = 0 & \text{in } Y_{\rm m} \\ \left[\boldsymbol{a} : \boldsymbol{e}_y\left(\boldsymbol{\xi}^{kh}\right)\right] \cdot \boldsymbol{n} = -[\boldsymbol{a} : \tilde{\boldsymbol{e}}] \cdot \boldsymbol{n} & \text{on } Y_{\rm fm} \end{cases} \tag{21}$$

where the tensor $\tilde{\boldsymbol{e}} = \tilde{e}_{ij} = \delta_{ik}\delta_{jh}$, $i, j = 1, 2, 3$, with $\delta_{ik}$ being the Kronecker symbol. In the same way, the vector $\zeta(y)$ satisfies

$$\begin{cases} \nabla_y \cdot [a : e_y(\zeta)] = 0 & \text{in } Y_m \\ [a : e_y(\zeta)] \cdot n = -I \cdot n & \text{on } Y_{fm} \end{cases} \tag{22}$$

Then, applying the second and third equations of (13) and (20), we obtain

$$\sigma_m^{(0)} = -[\alpha - a : e_y(\zeta) \cdot (I - \alpha)] p^{(0)} + a : (I + e_y(\xi)) : e_x(u_s^{(0)}) \tag{23}$$

Now, we define the total stress as

$$\sigma_T^{(0)} = \begin{cases} \sigma_m^{(0)} & \text{in } Y_m \\ \sigma_f^{(0)} & \text{in } Y_f \end{cases} \tag{24}$$

And let us define the average operator $\langle \cdot \rangle$

$$\langle \psi \rangle^l = \frac{1}{|Y|} \int_{Y_l} \psi(y) \, dy, \quad l = m, f \tag{25}$$

Averaging the total stress $\sigma_T^{(0)}$ over Y and using the first equation of (15), (16) and (23), we obtain

$$\langle \sigma_T^{(0)} \rangle = \langle \sigma_f^{(0)} \rangle^f + \langle \sigma_m^{(0)} \rangle^m = -\alpha_{eff} p^{(0)} + a_{eff} : e_x(u_s^{(0)}) \tag{26}$$

where

$$\begin{cases} \alpha_{eff} = \phi_m \alpha + \phi_c I - \langle a : e_y(\zeta) \rangle^m \cdot (I - \alpha) \\ a_{eff} = \langle a : [I + e_y(\xi)] \rangle^m \end{cases} \tag{27}$$

Integration of the seventh equation of (13) over $Y_m$, and applying the divergence theorem, the periodicity condition, and the fourth equation of (17), we obtain

$$\int_{Y_m} \nabla_x \cdot \sigma_m^{(0)} dy - \int_{Y_{fm}} n \cdot \sigma_f^{(1)} dA + \int_{Y_m} \rho_m g \, dy = 0 \tag{28}$$

Again, applying the divergence theorem, the periodicity condition, and integrating the second equation of (15), we see that

$$\int_{Y_f} \nabla_x \cdot \sigma_f^{(0)} dy + \int_{Y_{fm}} n \cdot \sigma_f^{(1)} dA + \int_{Y_f} \rho_f g \, dy = 0 \tag{29}$$

By adding the equations (28) and (29), we obtain

$$\int_Y \nabla_x \cdot \sigma_T^{(0)} dy + \int_Y \rho g \, dy = 0 \tag{30}$$

where, the mass density of the bulk material is defined as

$$\rho = \begin{cases} \rho_f & \text{in } Y_f \\ \rho_s & \text{in } Y_s \end{cases} \tag{31}$$

Then, by dividing by volume or area $|Y|$, (30) implies that

$$\nabla_x \cdot \langle \sigma_T^{(0)} \rangle + \langle \rho \rangle g = 0 \tag{32}$$

In the derivation of (32), a volume averaging theorem (Whitaker, 1999) has been used together with the periodicity condition, which implies that

$$\left\langle \nabla_x \cdot \boldsymbol{\sigma}_T^{(0)} \right\rangle = \nabla_x \cdot \left\langle \boldsymbol{\sigma}_T^{(0)} \right\rangle + \frac{1}{|Y|}\int_{Y_{fm}} \boldsymbol{n} \cdot \boldsymbol{\sigma}_T^{(0)} dA + \frac{1}{|Y|}\int_{Y_{pb}} \boldsymbol{n} \cdot \boldsymbol{\sigma}_T^{(0)} dA = \nabla_x \cdot \left\langle \boldsymbol{\sigma}_T^{(0)} \right\rangle \tag{33}$$

The equation (32) is the macroscopic Navier equation for a deformable vuggy porous medium.

### 3.4 The analysis of fluid flow behavior

In this section, it follows from the sixth and eighth equations of (13), the second, third, fourth equations of (15), and the first, sixth, seventh equations of (17) that

$$\begin{cases} \mu \boldsymbol{k}^{-1} \boldsymbol{v}_m^{(0)} + \nabla_y p_p^{(1)} + \nabla_x p_p^{(0)} - \rho_f \boldsymbol{g} = 0 & \text{in } Y_m \\ \nabla_y \cdot \boldsymbol{v}_m^0 = 0 & \text{in } Y_m \\ -2\mu \nabla_y \cdot \boldsymbol{e}\left(\boldsymbol{v}_f^{(0)}\right) + \nabla_y p_f^{(1)} + \nabla_x p_f^{(0)} - \rho_f \boldsymbol{g} = 0 & \text{in } Y_f \\ \nabla_y \cdot \boldsymbol{v}_f^{(0)} = 0 & \text{in } Y_f \\ \boldsymbol{v}_f^{(0)} \cdot \boldsymbol{n} = \left(\boldsymbol{v}_m^{(0)} + \dot{\boldsymbol{u}}_s^{(0)}\right) \cdot \boldsymbol{n} & \text{on } Y_{fm} \\ \boldsymbol{n} \cdot 2\mu \boldsymbol{e}_y\left(\boldsymbol{v}_f^{(0)}\right) \cdot \boldsymbol{n} = p_f^{(1)} - p_p^{(1)} & \text{on } Y_{fm} \\ \boldsymbol{\tau} \cdot 2\mu \boldsymbol{e}_y\left(\boldsymbol{v}_f^{(0)}\right) \cdot \boldsymbol{n} = -\frac{\beta}{\sqrt{\boldsymbol{\tau} \cdot \boldsymbol{k} \cdot \boldsymbol{\tau}}} \left(\boldsymbol{v}_f^{(0)} - \dot{\boldsymbol{u}}_s^{(0)}\right) \cdot \boldsymbol{\tau} & \text{on } Y_{fm} \end{cases} \tag{34}$$

Applying the equations (14) and (18) leads to

$$\begin{cases} \mu \boldsymbol{k}^{-1} \boldsymbol{v}_m^{(0)} + \nabla p_p^{(1)} = -\left(\nabla p^{(0)} - \rho_f \boldsymbol{g}\right) & \text{in } Y_m \\ \nabla_y \cdot \boldsymbol{v}_m^0 = 0 & \text{in } Y_m \\ -2\mu \nabla_y \cdot \boldsymbol{e}_y\left(\boldsymbol{v}_f^{(*)}\right) + \nabla_y p_f^{(1)} = -\left(\nabla p^{(0)} - \rho_f \boldsymbol{g}\right) & \text{in } Y_f \\ \nabla_y \cdot \boldsymbol{v}_f^{(*)} = 0 & \text{in } Y_f \\ \boldsymbol{v}_f^{(*)} \cdot \boldsymbol{n} = \boldsymbol{v}_m^{(0)} \cdot \boldsymbol{n} & \text{on } Y_{fm} \\ \boldsymbol{n} \cdot 2\mu \boldsymbol{e}_y\left(\boldsymbol{v}_f^{(*)}\right) \cdot \boldsymbol{n} = p_f^{(1)} - p_p^{(1)} & \text{on } Y_{fm} \\ \boldsymbol{\tau} \cdot 2\mu \boldsymbol{e}_y\left(\boldsymbol{v}_f^{(*)}\right) \cdot \boldsymbol{n} = -\frac{\beta}{\sqrt{\boldsymbol{\tau} \cdot \boldsymbol{k} \cdot \boldsymbol{\tau}}} \boldsymbol{v}_f^{(*)} \cdot \boldsymbol{\tau} & \text{on } Y_{fm} \end{cases} \tag{35}$$

where

$$\boldsymbol{v}_f^{(*)} = \boldsymbol{v}_f^{(0)} - \dot{\boldsymbol{u}}_s^{(0)} \tag{36}$$

The first and third equations of (35) imply that

$$\begin{cases} p_p^{(1)}(\boldsymbol{x},\boldsymbol{y}) = \pi_p^i(\boldsymbol{y})\left(\rho_f g_i - \partial_i p^{(0)}\right)(\boldsymbol{x},t) \\ p_f^{(1)}(\boldsymbol{x},\boldsymbol{y}) = \pi_f^i(\boldsymbol{y})\left(\rho_f g_i - \partial_i p^{(0)}\right)(\boldsymbol{x},t) \\ \boldsymbol{v}_m^{(0)}(\boldsymbol{x},\boldsymbol{y},t) = \frac{1}{\mu}\omega_m^i(\boldsymbol{y})\left(\rho_f g_i - \partial_i p^{(0)}\right)(\boldsymbol{x},t) \\ \boldsymbol{v}_f^{(*)}(\boldsymbol{x},\boldsymbol{y},t) = \frac{1}{\mu}\omega_f^i(\boldsymbol{y})\left(\rho_f g_i - \partial_i p^{(0)}\right)(\boldsymbol{x},t) \end{cases} \tag{37}$$

for each $i$-th dimension ($i=1,\ldots,d$). With $e^i$ being the standard Cartesian basis vector in the $i$-th direction, let $\pi_l^i(y)$ and $\omega_l^i(y)$ ($l$=f, m) be the periodic solution of the following Darcy-Stokes problem in the base cell

$$\begin{cases} \mu \boldsymbol{k}^{-1}\boldsymbol{\omega}_m^i + \nabla \pi_m^i = \boldsymbol{e}^i & \text{in } Y_m \\ \nabla_y \cdot \boldsymbol{\omega}_m^i = 0 & \text{in } Y_m \\ -2\mu\Delta_y \boldsymbol{\omega}_f^i + \nabla_y \pi_f^i = \boldsymbol{e}^i & \text{in } Y_f \\ \nabla_y \cdot \boldsymbol{\omega}_f^i = 0 & \text{in } Y_f \\ \boldsymbol{\omega}_f^i \cdot \boldsymbol{n} = \boldsymbol{\omega}_m^i \cdot \boldsymbol{n} & \text{on } Y_{fm} \\ \boldsymbol{n} \cdot 2\mu e_y(\boldsymbol{\omega}_f^i) \cdot \boldsymbol{n} = \pi_f^i - \pi_m^i & \text{on } Y_{fm} \\ \boldsymbol{\tau} \cdot 2\mu e_y(\boldsymbol{\omega}_f^i) \cdot \boldsymbol{n} = -\dfrac{\beta}{\sqrt{\boldsymbol{\tau} \cdot \boldsymbol{k} \cdot \boldsymbol{\tau}}}\boldsymbol{\omega}_f^i \cdot \boldsymbol{\tau} & \text{on } Y_{fm} \end{cases} \quad (38)$$

We now apply the averaging operator in the following sense

$$\left\langle \boldsymbol{v}_T^{(0)} \right\rangle = \left\langle \boldsymbol{v}_m^{(0)} \right\rangle^m + \left\langle \boldsymbol{v}_f^{(*)} \right\rangle = -\sum_i \frac{1}{\mu|Y|}\left( \int_{Y_m} \boldsymbol{\omega}_m^i \mathrm{d}y + \int_{Y_f} \boldsymbol{\omega}_f^i \mathrm{d}y \right)\left( \partial_i p^{(0)} - \rho_f g_i \right) \quad (39)$$

Then together with the fourth equation of (13) and (36), we see that

$$\begin{aligned}\left\langle \boldsymbol{v}_T^{(0)} \right\rangle &= \left\langle \boldsymbol{v}_m^{(0)} \right\rangle^m + \left\langle \boldsymbol{v}_f^{(*)} \right\rangle^f = \left\langle \boldsymbol{v}_p^{(0)} - \phi \dot{\boldsymbol{u}}_s^{(0)} \right\rangle^m + \left\langle \boldsymbol{v}_f^{(0)} - \dot{\boldsymbol{u}}_s^{(0)} \right\rangle^f \\ &= \underbrace{\left\langle \boldsymbol{v}_p^{(0)} \right\rangle^m + \left\langle \boldsymbol{v}_f^{(0)} \right\rangle^f}_{\text{total pore fluid velocity}} - \underbrace{(\phi_c + \phi\phi_m)}_{\text{total porosity}}\dot{\boldsymbol{u}}_s^{(0)} = \left\langle \boldsymbol{v}_T^{(0)} \right\rangle^P - \phi_T \dot{\boldsymbol{u}}_s^{(0)} \end{aligned} \quad (40)$$

where, $\left\langle \boldsymbol{v}_T^{(0)} \right\rangle^P$ is the fluid velocity through the total pores in fractured or vuggy porous media, including the matrix pores and fractures (or vugs); $\phi_T$ is the total porosity of the fractured or vuggy porous media, including double porosity, i.e. matrix porosity and vuggs' porosity. Now let the matrix $k_\mathrm{eff}$ be defined by

$$k_\mathrm{eff}^{ij} = \frac{1}{|Y|}\left( \int_{Y_m} (\boldsymbol{\omega}_m^i)_j \mathrm{d}y + \int_{Y_f} (\boldsymbol{\omega}_f^i)_j \mathrm{d}y \right) \quad (41)$$

Then we can obtain the following macroscopic Darcy's equation

$$\begin{cases} \mu k_\mathrm{eff}^{-1}\left\langle \boldsymbol{v}_T^{(0)} \right\rangle + \nabla_x p^{(0)} - \rho_f \boldsymbol{g} = 0 \\ \left\langle \boldsymbol{v}_T^{(0)} \right\rangle = \left\langle \boldsymbol{v}_T^{(0)} \right\rangle^P - \phi_T \dot{\boldsymbol{u}}_s^{(0)} = -\dfrac{k_\mathrm{eff}}{\mu}\left( \nabla_x p^{(0)} - \rho_f \boldsymbol{g} \right) \end{cases} \quad (42)$$

## 3.5 The analysis of hydro-mechanical coupling equation

Finally, let us collect the ninth equation of (13) and the fifth equation of (15), which leads to

$$\begin{cases} \nabla_x \cdot \boldsymbol{v}_m^{(0)} + \nabla_y \cdot \boldsymbol{v}_m^{(1)} = -\boldsymbol{\alpha} : \left[ e_x(\dot{\boldsymbol{u}}_s^{(0)}) + e_y(\dot{\boldsymbol{u}}_s^{(1)}) \right] - \gamma \dot{p}^{(0)} & \text{in } Y_m \\ \nabla_x \cdot \boldsymbol{v}_f^{(0)} + \nabla_y \cdot \boldsymbol{v}_f^{(1)} = 0 & \text{in } Y_f \end{cases} \quad (43)$$

Following the definition of total Darcy velocity in (40) and (36), we require the following forms

$$\begin{cases} \nabla_x \cdot \boldsymbol{v}_m^{(0)} + \nabla_y \cdot \boldsymbol{v}_m^{(1)} = -\boldsymbol{\alpha} : \left[ \boldsymbol{e}_x \left( \dot{\boldsymbol{u}}_s^{(0)} \right) + \boldsymbol{e}_y \left( \dot{\boldsymbol{u}}_s^{(1)} \right) \right] - \gamma \dot{p}_p^{(0)} & \text{in } Y_m \\ \nabla_x \cdot \boldsymbol{v}_f^{(*)} + \nabla_y \cdot \boldsymbol{v}_f^{(1)} = -\boldsymbol{I} : \boldsymbol{e}_x \left( \dot{\boldsymbol{u}}_s^{(0)} \right) & \text{in } Y_f \end{cases} \quad (44)$$

where, the relation $\nabla_x \cdot \dot{\boldsymbol{u}}_s^{(0)} = \boldsymbol{I} : \boldsymbol{e}_x \left( \dot{\boldsymbol{u}}_s^{(0)} \right)$ has been used. Now integrating the above equations over $Y$, and applying the average operators, we find

$$\nabla_x \cdot \left\langle \boldsymbol{v}_T^{(0)} \right\rangle + \frac{1}{|Y|} \int_{Y_{fm}} \boldsymbol{n} \cdot \dot{\boldsymbol{u}}_s^{(1)} \mathrm{d}y = -\left\langle \boldsymbol{\alpha} : \left[ \boldsymbol{e}_x \left( \dot{\boldsymbol{u}}_s^{(0)} \right) + \boldsymbol{e}_y \left( \dot{\boldsymbol{u}}_s^{(1)} \right) \right] + \gamma \dot{p}^{(0)} \right\rangle^m - \left\langle \boldsymbol{I} : \boldsymbol{e}_x \left( \dot{\boldsymbol{u}}_s^{(0)} \right) \right\rangle^f \quad (45)$$

Then it follows from a further application of the divergence theorem and periodicity condition for the second term on the left hand side, and use of (23) that

$$\begin{aligned} \nabla_x \cdot \left\langle \boldsymbol{v}_T^{(0)} \right\rangle &= -\left( \phi_m \boldsymbol{\alpha} + \phi_c \boldsymbol{I} - (\boldsymbol{I} - \boldsymbol{\alpha}) : \left\langle \boldsymbol{e}_y (\boldsymbol{\xi}) \right\rangle^m \right) : \boldsymbol{e}_x \left( \dot{\boldsymbol{u}}_s^{(0)} \right) \\ &\quad - \left( \phi_m \gamma - (\boldsymbol{I} - \boldsymbol{\alpha}) : \left\langle \boldsymbol{e}_y (\boldsymbol{\zeta}) \right\rangle^m \cdot (\boldsymbol{I} - \boldsymbol{\alpha}) \right) \dot{p}^{(0)} \\ &= -\underbrace{\left( \phi_m \boldsymbol{\alpha} + \phi_c \boldsymbol{I} - \left\langle \boldsymbol{a} : \boldsymbol{e}_y (\boldsymbol{\zeta}) \right\rangle^m \cdot (\boldsymbol{I} - \boldsymbol{\alpha}) \right)}_{\boldsymbol{a}_{eff}} : \boldsymbol{e}_x \left( \dot{\boldsymbol{u}}_s^{(0)} \right) \\ &\quad - \underbrace{\left( \phi_m \gamma - (\boldsymbol{I} - \boldsymbol{\alpha}) : \left\langle \boldsymbol{e}_y (\boldsymbol{\zeta}) \right\rangle^m \cdot (\boldsymbol{I} - \boldsymbol{\alpha}) \right)}_{\gamma_{eff}} \dot{p}^{(0)} \end{aligned} \quad (46)$$

here we used the following relations (Chen et al., 2004)

$$\left\langle \boldsymbol{I} : \boldsymbol{e}_y (\boldsymbol{\xi}) \right\rangle^m = \left\langle \nabla_y \cdot \boldsymbol{\xi} \right\rangle^m = \left\langle \boldsymbol{a} : \boldsymbol{e}_y (\boldsymbol{\zeta}) \right\rangle^m \quad (47)$$

## 3.6 The macroscopic equations

With this, the macroscopic equations of a deformable vuggy porous medium are given by (26), (32), (42) and (46). And the macroscopic equations can be rewritten as follows:

$$\begin{cases} \nabla_x \cdot \left\langle \boldsymbol{\sigma}_T^{(0)} \right\rangle + \left\langle \rho \right\rangle \boldsymbol{g} = 0 \\ \left\langle \boldsymbol{\sigma}_T^{(0)} \right\rangle = -\boldsymbol{\alpha}_{eff} p^{(0)} + \boldsymbol{a}_{eff} : \boldsymbol{e}_x \left( \boldsymbol{u}_s^{(0)} \right) \\ \left\langle \boldsymbol{v}_T^{(0)} \right\rangle = \left\langle \boldsymbol{v}_T^{(0)} \right\rangle^p - \phi_T \dot{\boldsymbol{u}}_s^{(0)} = -\frac{\boldsymbol{k}_{eff}}{\mu} \left( \nabla_x p^{(0)} - \rho_f \boldsymbol{g} \right) \\ \nabla_x \cdot \left( \left\langle \boldsymbol{v}^{(0)} \right\rangle^f - \phi \dot{\boldsymbol{u}}_s^{(0)} \right) = -\boldsymbol{\alpha}_{eff} : \boldsymbol{e}_x \left( \dot{\boldsymbol{u}}_s^{(0)} \right) - \gamma_{eff} \dot{p}^{(0)} \end{cases} \quad (48)$$

where the corresponding coefficients are defined as following:

$$\begin{cases} \boldsymbol{k}_{eff} \to k_{eff}^{ij} = \left\langle \left( \omega_m^i \right)_j \right\rangle^m + \left\langle \left( \omega_f^i \right)_j \right\rangle^f \\ \boldsymbol{a}_{eff} \to \left( a_{eff} \right)_{ijkh} = \left\langle a_{ijkh} + a_{ijlm} e_{ylm} \left( \boldsymbol{\xi}^{kh} \right) \right\rangle^m \\ \boldsymbol{\alpha}_{eff} \to \left( \alpha_{eff} \right)_{kh} = \phi_m \alpha_{kh} + \phi_c I_{kh} - \left( I_{ij} - \alpha_{ij} \right) \left\langle e_{yij} \left( \boldsymbol{\xi}^{kh} \right) \right\rangle^m \\ \gamma_{eff} = \phi_m \gamma - \left( I_{ij} - \alpha_{ij} \right) \left\langle e_{yjk} (\boldsymbol{\zeta}) \right\rangle^m \left( I_{ki} - \alpha_{ki} \right) \end{cases} \quad (49)$$

The formula of (48) are classical Biot's equations. That is, the macroscopic hydro-mechanical behavior of vuggy porous media is also described by using the Biot's theory in the framework of small strains. However, the corresponding coefficients are more complex and difficult to evaluate. Here, we summarize the properties of the above coefficients. At first, it can be shown that $\boldsymbol{k}_{\text{eff}}$ is positive definite and symmetric (Arbogast and Lehr, 2006). We also note that $\boldsymbol{\alpha}_{\text{eff}}$ is symmetric. Second, it is easy to prove that the effective elasticity coefficient $\boldsymbol{a}_{\text{eff}}$ is symmetric and positive definite (Chen et al., 2004; Auriault et al., 2009). Finally, the last equation of (49) shows that $\gamma_{\text{eff}} > 0$.

## 4. Numerical experiments and discussions

In this section, several numerical experiments are conducted to verify the validation of the proposed macroscopic equations and to evaluate the macroscopic hydro-mechanical properties. The identification and the use of the macroscopic constitutive equations of the vuggy porous media require the solution of the cell problems (21)-(22) for the mechanical part and cell problem (38) for the fluid flow part. However, even for simple configurations of the elementary cell, those problems cannot be solved analytically and a numerical computation is needed. In this work, all computations are performed by using the finite element method (FEM). In this work, the cell problems of the mechanical part (Eqs. 21 and 22) are solved using classical Galerkin FEM. For the cell problem of the fluid part (Eq 38), we use a mixed FEM in the porous flow and free fluid flow regions. For more details on the said numerical schemes, the reader is referred to (Bower, 2009; Arbogast and Brunson, 2007; Huang et al., 2011).

### 4.1 A porous medium with a circular vug

As shown in Figure 4-a, a periodic medium ($10^{-3}$ m × $10^{-3}$ m) with a circular vug is considered, and the porous matrix is assumed to be isotropic and homogeneous (i.e. no vug). We study the impacts of Poisson's ratio of matrix and volumetric fraction of the vug on effective Biot coefficient $\alpha_{\text{eff}}$ with two cases. In Case 1, the matrix is assumed to be non-porous, which is the same as the work of (Lydzba and Shao, 2000), and the Young modulus and Biot coefficient $\boldsymbol{\alpha}$ of matrix are 1 GPa and $\boldsymbol{0}$, respectively. In Case 2, the matrix is porous, and the parameters are given as: Young modulus $E$, 1 GPa; the Poisson's ratio $v$, 0.3; Porosity of matrix $\phi$, 0.126; Biot coefficient $\boldsymbol{\alpha}$, 0.334$\boldsymbol{I}$; Storage coefficient $\gamma$, 0.218 GPa$^{-1}$. Due to the symmetries of the porous medium, it yields an isotropic macroscopic model. Herein, we applied the plane strain numerical simulation.

At first, we keep the volumetric fraction of vug as a constant (i.e. $\phi_c$=0.126), and change Poisson's ratio of the matrix to different values. Then keeping the Poisson's ratio as a constant ($v$=0.2), we change the volumetric fraction of vug to different values. Figure 4-a and 4-b show the solutions of the cell base problems, i.e. Eqs. (21)-(22). The $\alpha_{\text{eff}}$ obtained for different situations are plotted in Figure 5, and excellent agreements between the current work for non-porous matrix and the references (Lydzba and Shao, 2000) are achieved. Comparing to the non-porous matrix case, the Biot coefficient $\alpha_{\text{eff}}$ of the porous matrix cases is increased, which means the hydro-mechanical coupling behavior becomes stronger. At the same time, this enhancement feature becomes less marked with the increase of the Poisson's ratio and volumetric fraction of the vug. Table 1 shows the comparison of the calculated effective Biot's coefficients between this work and the references.

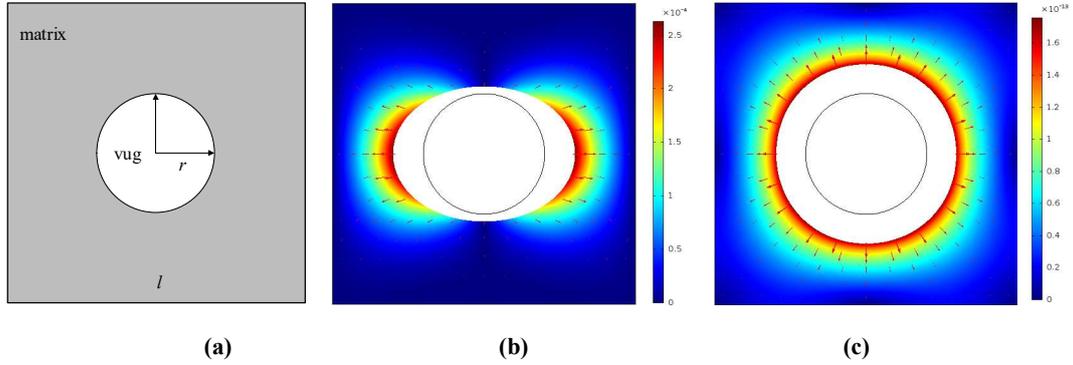

|  (a)  |  (b)  |  (c)  |

**Figure 4: (a) Schematic of a porous medium with a circular vug, (b) the solution $\xi^{11}$ of base problem (21) corresponding to the $x$-direction displacement, (c) the solution $\zeta$ of base problem (22) corresponding to the total displacement results.**

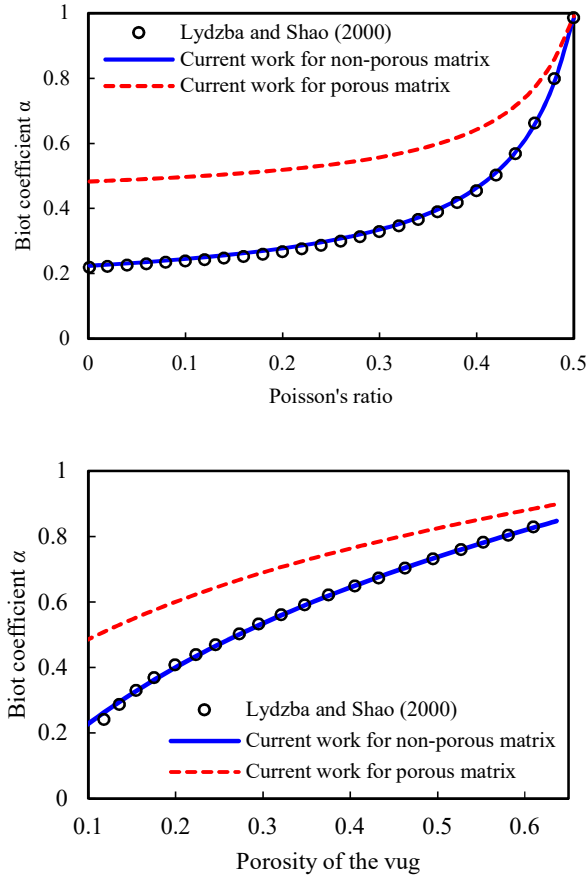

**Figure 5: Biot coefficient $\alpha_{11}$ versus Poisson's ratio (top) and porosity of the vug (bottom).**

To assess the validity of this work for calculating effective permeability, a same porous medium as above with different vug radius is considered, and the matrix permeability is 10 mD (milli Darcy). The isotropic effective permeability for different vug radius is presented in Table 1, and results of the current model are compared with those of (Huang et al., 2011) and (Wan and Eghbalian, 2016b). As shown in Table 2, the effective permeability increases obviously with the increasing of vug's radius, and our numerical results get a good match with that of the reference results. It should be noted that the results of (Wan and Eghbalian, 2016b) are little different from the current work and (Huang et al., 2011). This is because the former used the continuity condition of tangential velocity but the last two used the BJS slip conditions. The impacts of the BJS slip coefficient on the effective permeability have been investigated in our previous work (Huang et al., 2011).

Table 1: The comparison of the effective Biot's coefficients between this work and the references

| Biot's Coefficients | Lewandowska and Auriault (2013) | Wan and Eghbalian, (2016b) | This work |
|---|---|---|---|
| $\alpha_{eff}$ | 0.556 | 0.517 | 0.55715 |
| $\gamma_{eff}$ (GPa$^{-1}$) | 0.287 | - | 0.28697 |

Table 2: The effective permeability of base cell with different vug radius

| 2r/L | $k_{11} = k_{22}$ (mD) | | |
|---|---|---|---|
| | Wan and Eghbalian (2016b) | Huang et al. (2011) | This work |
| 0.02 | 1.00063E+01 | 1.000628E+01 | 1.000628E+01 |
| 0.1 | 1.01571E+01 | 1.015803E+01 | 1.015810E+01 |
| 0.2 | 1.06283E+01 | 1.064756E+01 | 1.064840E+01 |
| 0.4 | 1.25133E+01 | 1.287336E+01 | 1.287413E+01 |
| 0.6 | — | 1.790391E+01 | 1.790512E+01 |
| 0.8 | — | 3.104431E+01 | 3.104633E+01 |
| 0.98 | — | 1.374829E+02 | 1.374740E+02 |

## 4.2 A porous medium with a penny-shaped vug

In this subsection, a porous medium with a penny-shaped vug is used to verify the current work. The semi-axes of the vug are 0.2 mm in the *x*-direction and 0.002 mm in the *y*-direction, and the parameters are the same as Case 1 in section 4.1. The effective Biot's coefficients are reported by (Lewandowska and Auriault, 2013; Wan and Eghbalian, 2016b) are compared to this work, which is shown in Table 3. It can be seen that the results of current work are in excellent agreement with those of (Lewandowska and Auriault, 2013), while there is a slight error between those of (Wan and Eghbalian, 2016b). It should be note that the macroscopic model proposed by Wan and Eghbalian is aimed to fractured porous media. When the shape of vugs becomes closer, the closed-form of the poroelastic parameters will agree better with our work.

Two calculations were performed to solve problems (21) by applying a macroscopic unit strain along the *x*-direction and *y*-direction, respectively. Displacements $\xi^{11}$ and $\xi^{22}$ are presented in Figure 6. It should be noted that there are small differences between our current work and Lewandowska and Auriault (2013). This is because they used the simplified boundary conditions for the base problem (21) as described in section 1. Although the Biot coefficient $\alpha$ of the porous matrix is isotropic, the effective Biot coefficient of the porous medium is anisotropic due to the penny-shaped vug.

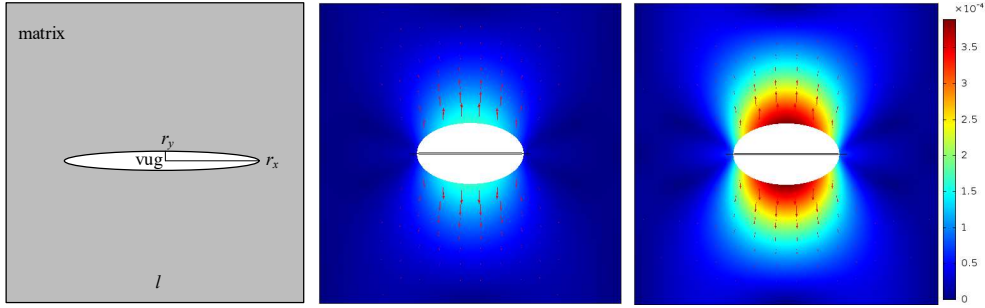

**Figure 6: (a) Schematic of a porous medium with a penny-shaped vug, (b) the solution $\xi^{11}$ of base problem (21) corresponding to the total displacement, (c) the solution $\xi^{22}$ of base problem (21) corresponding to the total displacement results.**

**Table 3: The comparison of the effective Biot's coefficients between this work and the references**

| Biot's Coefficients | Lewandowska and Auriault (2013) | Wan and Eghbalian, (2016b) | This work |
|---|---|---|---|
| $\alpha_{\text{eff}}^{11}$ ($\xi^{11}$) | 0.404 | 0.394 | 0.4043 |
| $\alpha_{\text{eff}}^{22}$ ($\xi^{22}$) | 0.497 | 0.490 | 0.4975 |
| $\gamma_{\text{eff}}$ (GPa$^{-1}$) | 0.298 | - | 0.2981 |

The effective elasticity tensor of the vuggy porous medium is compared to that of the porous matrix, as listed below. All the components of effective elastic tensor are decreased due to the presence of a penny-shaped vug, especially in the *y*-direction (i.e. along the semi-minor axis).

$$\boldsymbol{a} = [a]_{mn} = \begin{bmatrix} 1.3462 & 0.5769 & 0 \\ 0.5769 & 1.3462 & 0 \\ 0 & 0 & 0.3846 \end{bmatrix} \times 10^9, \quad \boldsymbol{a}_{\text{eff}} = [a_{\text{eff}}]_{mn} = \begin{bmatrix} 1.1166 & 0.3638 & 0 \\ 0.3638 & 0.8473 & 0 \\ 0 & 0 & 0.3024 \end{bmatrix} \times 10^9$$

## 4.3 The effects of vugs' geometric parameters

In this subsection, the evaluations of the geometric dependences of effective poroelastic parameters are investigated (as shown in Figure 7). The sizes and parameters of the periodic media are the same as the Case 2 in section 4.1, but with Poisson's ratio 0.25. In addition, the volume fraction of vugs is 0.126 in all examples. As shown in Table 4, the numerical results indicate that the effective parameters are correlated with the vug's shape.

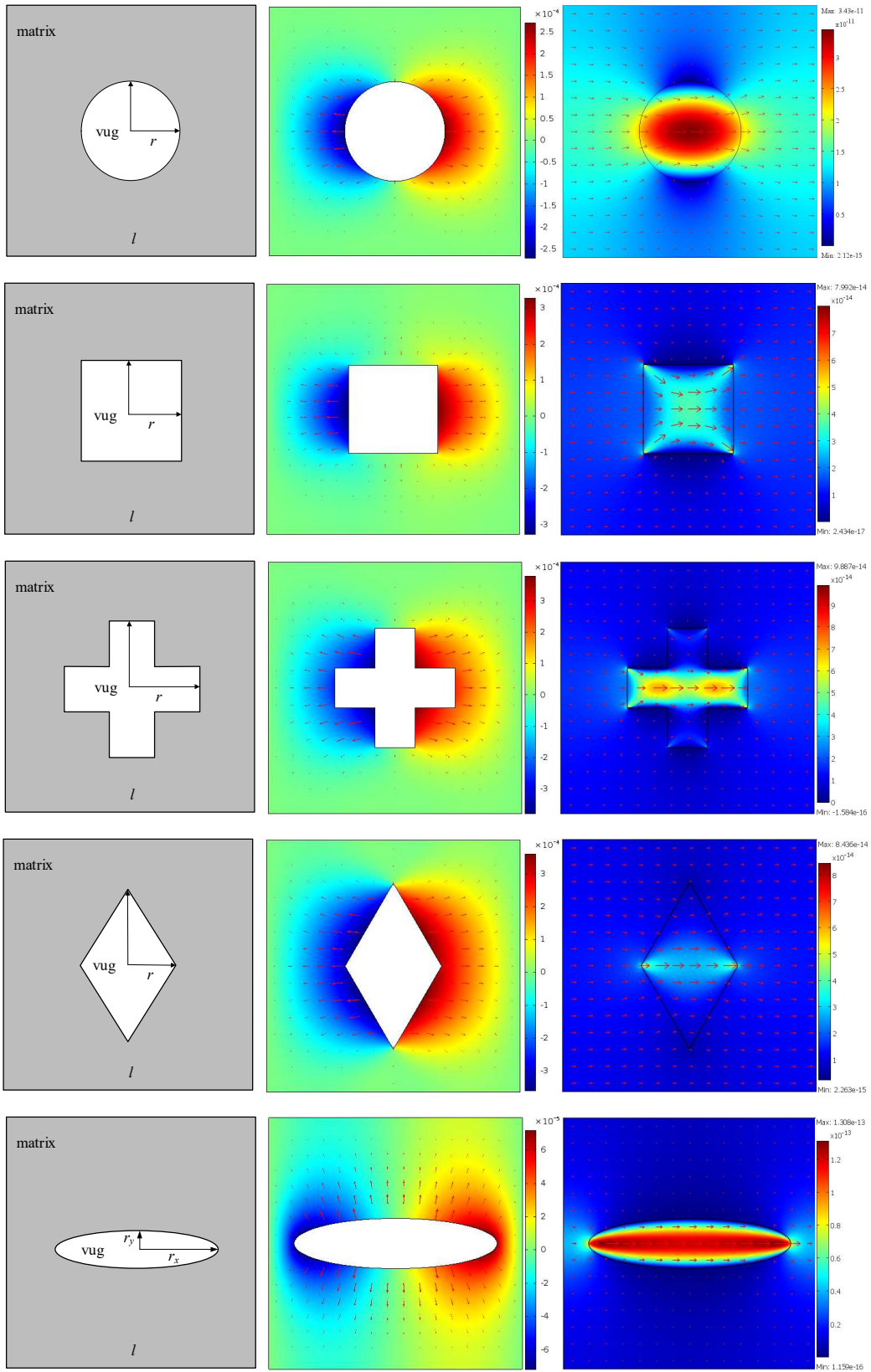

**Figure 7: Geometry of porous media with different shape vugs (left), the solution $\xi^{11}$ of base problem (21) corresponding to the *x*-direction displacement (middle), and the solution $\omega_l^i(y)$ of base problem (38)**

corresponding to the *x*-direction velocity profile (right).

Table 4: The comparison of the effective coefficients between different shape vugs

| Effective properties | matrix | circular | square | cross-shape | rhombus | ellipse |
|---|---|---|---|---|---|---|
| $\gamma_{\text{eff}}$ (1/GPa) | 0.218 | 0.28792 | 0.31138 | 0.33744 | 0.33811 | 0.38030 |
| $\alpha_{11}$ | 0.334 | 0.55850 | 0.59240 | 0.63000 | 0.68700 | 0.58670 |
| $\alpha_{22}$ | 0.334 | 0.55850 | 0.59240 | 0.63000 | 0.57490 | 0.79710 |
| $a_{1111}$ (GPa) | 1.2 | 0.82625 | 0.79061 | 0.70950 | 0.56220 | 0.92757 |
| $a_{2222}$ (GPa) | 1.2 | 0.82625 | 0.79061 | 0.70950 | 0.86512 | 0.35898 |
| $a_{1212}$ (GPa) | 0.4 | 0.26379 | 0.22629 | 0.23379 | 0.25592 | 0.18956 |
| $a_{1122}$ (GPa) | 0.4 | 0.23437 | 0.18863 | 0.17935 | 0.15603 | 0.12851 |
| $k_{11}$ ($10^{-14}$ m$^2$) | 1.0 | 1.287338 | 1.315135 | 1.382383 | 1.226424 | 2.046991 |

For effective Darcy permeability $k_{\text{eff}}$, its directional value is dependent on the length of a vug along with the corresponding direction. And it increases with the increasing length. This implies that the vug connectivity may be the most important factor for effective permeability. For effective elastic Young's modulus, the second equation of (49) indicates that its value is dependent on the averaging stress under the traction of $\xi^{kh}$ imposed at porous-fluid interface. Therefore, its value is strongly correlated to the interface distribution and position on the medium. The numerical results of $a_{ijkl}$ in Table 4 confirm that.

For the effective Biot's coefficient, the third equation of (49) indicates that its value depends on the porosity of vugs and the traction of $\xi^{kh}$ imposed at porous-fluid interface. In this example, the porosity of vugs is same. As a result, the effective Biot's coefficient also relies on the shape and distribution of the porous-fluid interface. The parameter $\gamma_{\text{eff}}$ has the similar variation trend to the effective permeability. Actually, the shape of vug determines the interface between porous matrix and free fluid flow region, which is the key for the analysis.

## 4.4 Analysis of a 3D vuggy porous medium

Vug are defined as visible pores that are significantly larger than adjacent grains or crystals (Lucia, 2007). They may be formed through a variety of processes. Most commonly, vugs may form when mineral crystals or fossils inside a rock matrix are later removed through erosion or dissolution processes, leaving behind irregular voids. In this subsection, we will give a simplified 3D model for this typical vug, as shown in Figure 8. There is a tortuous pipe, and the medium part was enlarged due to the erosion process. The volume fraction of the vug is 0.0791.

Figure 9 shows the meshes for different cell problems, which are corresponding to solid part and fluid flow part. The numerical results of different components have been shown in Figure 10. The calculated effective elastic parameters and the effective Darcy permeability are listed in Table 5. As illustrated in Figure 8, the tortuous vug actually connect with each other along the *x*-direction due to the periodicity. As a result, the component of the effective permeability in *x*-direction is larger than the components in *y* and *z*-directions will decrease more.

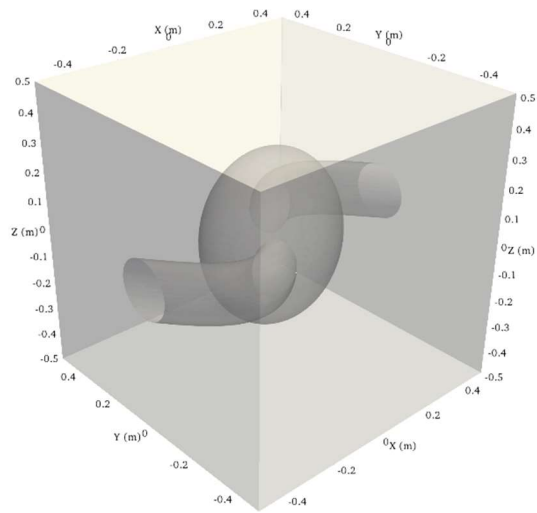

**Figure 8: 3D Geometry of a porous medium with a tortuous vug**

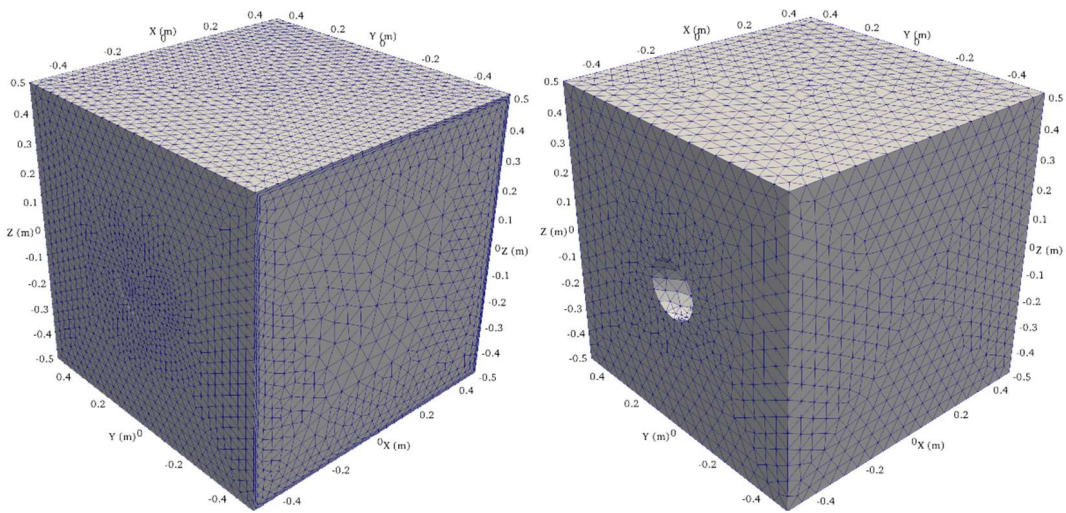

**Figure 9: 3D Geometry of a porous medium with a tortuous vug**

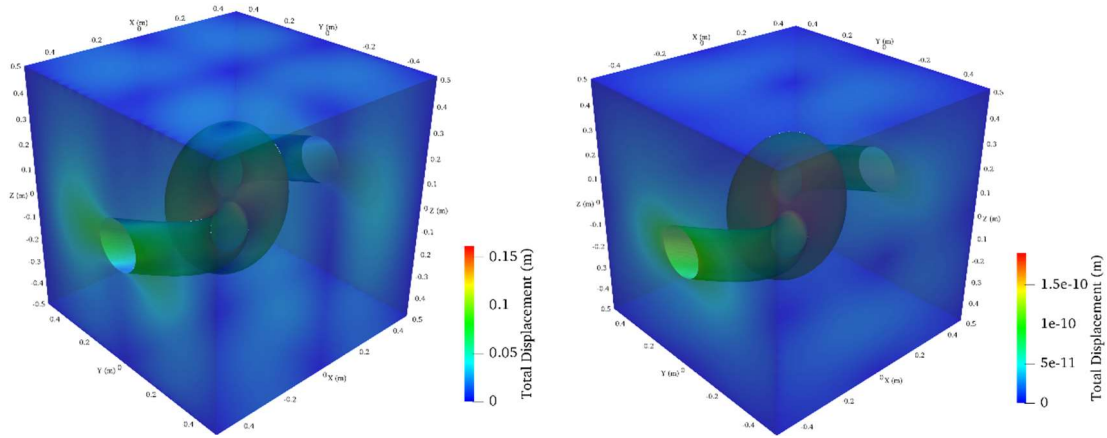

(a) the solution $\xi^{12}$ of cell problem (21)     (b) the solution $\zeta$ of cell problem (22)

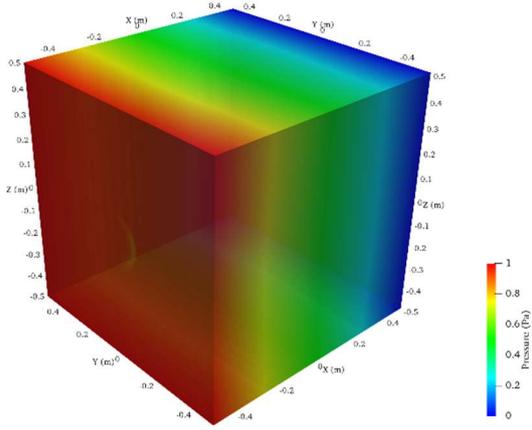 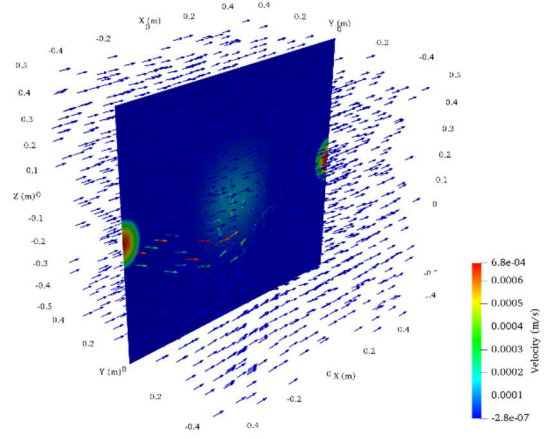

(c) the solution $\left(\pi_l^i\right)_{e_x}$ of cell problem (38)      (d) the solution $\left(\omega_l^i\right)_{e_x}$ of cell problem (38)

Figure 10: The numerical results of several components of the solutions of cell problems

Table 5: The calculated effective parameters based on the solutions of the cell problems

| Effective properties | Porous matrix | The 3D vuggy porous medium |
|---|---|---|
| $\gamma_{\text{eff}}$ (1/GPa) | 0.218 | 0.27237 |
| $\alpha_{11}$ | 0.334 | 0.46771 |
| $\alpha_{22}$ | 0.334 | 0.55850 |
| $k_{11}$ (m$^2$) | $1.0\times 10^{-14}$ | $1.0562\times 10^{-05}$ |
| $k_{22}$ (m$^2$) | $1.0\times 10^{-14}$ | $1.2536\times 10^{-14}$ |
| $k_{33}$ (m$^2$) | $1.0\times 10^{-14}$ | $1.2841\times 10^{-14}$ |

The calculated effective elasticity tensor is compared to that of the porous matrix, as listed below. All the components of effective elastic tensor are decreased due to the presence of the tortuous vug, especially along the $y$ and $z$-directions.

$$\boldsymbol{a}=[a]_{mn}=\begin{bmatrix} 1.3462 & 0.5769 & 0.5769 & 0 & 0 & 0 \\ 0.5769 & 1.3462 & 0.5769 & 0 & 0 & 0 \\ 0.5769 & 0.5769 & 1.3462 & 0 & 0 & 0 \\ 0 & 0 & 0 & 0.3846 & 0 & 0 \\ 0 & 0 & 0 & 0 & 0.3846 & 0 \\ 0 & 0 & 0 & 0 & 0 & 0.3846 \end{bmatrix}\times 10^9$$

$$\boldsymbol{a}_{\text{eff}}=[a_{\text{eff}}]_{mn}=\begin{bmatrix} 1.1219 & 0.4316 & 0.4446 & 0 & 0 & 0 \\ 0.4316 & 1.0517 & 0.4299 & 0 & 0 & 0 \\ 0.4446 & 0.4299 & 1.1121 & 0 & 0 & 0 \\ 0 & 0 & 0 & 0.3188 & 0 & 0 \\ 0 & 0 & 0 & 0 & 0.3157 & 0 \\ 0 & 0 & 0 & 0 & 0 & 0.3244 \end{bmatrix}\times 10^9$$

# 5. Conclusions

We have systematically derived the macroscopic equations for the mechanical behavior of a deformable vuggy porous medium saturated with a Newtonian fluid via the homogenization theory. The developed formulation presented in this paper is novel since it takes into account the general Beavers-Joseph-Saffman (BJS) interface boundary conditions for Biot-Stokes system. In the case of Biot-Stokes coupling hydro-mechanical system, these macroscopic equations coincide with classical Biot's equations. And the homogenization approach determines the form of the macroscopic constitutive relationships between variables, and shows how to compute the poroelastic coefficients in these relationships. Several numerical tests have been conducted to demonstrated the calculation procedures. The numerical results indicates that the existence of the vugs have significant impacts on the effective elastic and hydro parameters of the vuggy porous media.

In addition, we note that the calculations of the macroscopic properties only depend on the decoupling three cell problems, i.e., two Navier equations for elastic problem and one Darcy-Stokes flow problem with BJS condition. It also show that our homogenized results provide a natural way of modeling realistic deformable vuggy porous media. In this work, only small strain and elastic problem is considered. In the future, the finite deformation analysis will be conducted, and the corresponding elastoplastic problem is also the next topic.

# Acknowledgement

This work was supported by National Nature Science Foundation of China (Grant No. 51404292), the Fundamental Research Funds for the Central Universities (18CX05029A, 17CX06007), National Science and Technology Major Project (2017ZX05009001, 2016ZX05060-010).